\documentclass{PoS}

\title{Determining SUSY particle mixing with polarized hadron beams}

\ShortTitle{Determining SUSY particle mixing with polarized hadron beams}

\author{\speaker{Michael Klasen}%
         \\
        Laboratoire de Physique Subatomique et de Cosmologie,
        Universit\'e Joseph Fourier / CNRS-IN2P3 / INPG,
        53 Avenue des Martyrs, F-38026 Grenoble, France\\
        E-mail: \email{klasen@lpsc.in2p3.fr}

\vspace*{-95mm}
\noindent {\rm LPSC 10-052}\\
\vspace*{ 86mm}

}

\abstract{
While SUSY particles, if they exist at the TeV-scale, will be discovered
at the Tevatron or the LHC, the determination of the SUSY-breaking scenario and
its free parameters will require additional information, e.g.\ from a future
International Linear Collider. We point out that such information, in
particular on SUSY-particle mixing and the associated soft SUSY-breaking
parameters, can also be obtained from measurements at existing or future polarized
hadron colliders, since the polarization of initial-state quarks, transmitted
through weak gauge bosons or squarks, can be strongly correlated with the
helicity and gaugino/higgsino mixing of final-state sleptons, squarks, neutralinos
and charginos.
}

\FullConference{XVIII International Workshop on Deep-Inelastic Scattering and Related Subjects\\
		 April 19 -23, 2010\\
		 Convitto della Calza, Firenze, Italy}

\begin{document}

\def\d{{\rm d}}

\def\lp{\left. }
\def\rp{\right. }
\def\lr{\left( }
\def\rr{\right) }
\def\le{\left[ }
\def\re{\right] }
\def\lg{\left\{ }
\def\rg{\right\} }
\def\lb{\left| }
\def\rb{\right| }

\def\beq{\begin{equation}}
\def\eeq{\end{equation}}
\def\bea{\begin{eqnarray}}
\def\eea{\end{eqnarray}}

\section{Motivation}

Weak-scale supersymmetry (SUSY) continues to be one of the best-motivated
extensions of the Standard Model (SM) of particle physics, and the search
for SUSY particles is one of the top priorities at current high-energy hadron
colliders. While SUSY particles, if they exist at the TeV-scale,
will be discovered at the Tevatron or the LHC and their masses and some of
their properties will ultimately be measured with an accuracy of about 10\%
in the high-energy and high-luminosity phase of the LHC, it is today unanimously
accepted that the determination of the SUSY-breaking scenario as well as its
free parameters will require additional and more precise measurements, e.g.\
from a future International Linear Collider \cite{Weiglein:2004hn}. In
this article, we point out that additional information, in particular on
SUSY-particle mixing and the associated soft SUSY-breaking parameters, may
also be obtained from measurements at existing or future polarized hadron
colliders such as RHIC or the Tevatron and LHC after polarization upgrades
\cite{Bozzi:2004qq}.

\section{SUSY particle mixing}

Since the two SUSY partners of a chiral lepton or quark doublet are scalars
and carry identical quantum numbers, left- and right-handed sleptons $\tilde{l}_
{L,R}$ or squarks $\tilde{q}_{L,R}$ can mix into lighter and heavier mass
eigenstates $\tilde{l}_{1,2}$ and $\tilde{q}_{1,2}$. The mixing is proportional to
the off-diagonal elements of the sfermion mass matrix
\beq
 {\mathcal M}^2 =
 \lr\begin{array}{cc}
  m_{LL}^2+m_f^2  &
  m_f m_{LR}^\ast \\
  m_f m_{LR}      &
  m_{RR}^2+m_f^2
 \end{array}\rr
\eeq
with
\bea
 m_{LL}^2&=&(T_f^3-e_f\sin^2\theta_W)m_Z^2\cos2\beta+m_{\tilde{L}}^2,\\
 m_{RR}^2&=&e_f\sin^2\theta_W m_Z^2\cos2\beta+\left\{\begin{array}{l}
 m_{\tilde{U}}^2\hspace*{5mm}{\rm for~up-type~fermions},\\
 m_{\tilde{D}}^2\hspace*{5mm}{\rm for~down-type~sfermions},\end{array}\right.\\
 m_{LR}  &=&A_f-\mu^\ast\left\{\begin{array}{l}
 \cot\beta\hspace*{5mm}{\rm for~up-type~sfermions}\\
 \tan\beta\hspace*{5mm}{\rm for~down-type~sfermions}\end{array}\right.
\eea
and thus to the mass of the SM fermion $m_f$ and the soft SUSY-breaking term
$A_f$ of the trilinear Higgs-sfermion-sfermion interaction and the off-diagonal
Higgs mass parameter $\mu$ in the MSSM Lagrangian, divided or multiplied by $\tan
\beta=s_\beta/c_\beta=v_u/v_d$, the ratio of the vacuum expectation values
$v_{u,d}$ of the two Higgs doublets. This mass matrix is diagonalized by a unitary
matrix
\beq
 S = \lr \begin{array}{cc}~~\,\cos\theta_{\tilde{f}} &
                              \sin\theta_{\tilde{f}} \\
                             -\sin\theta_{\tilde{f}} &
                              \cos\theta_{\tilde{f}} \end{array} \rr
 \hspace*{3mm} {\rm with} \hspace*{3mm}
   \lr \begin{array}{c} \tilde{f}_1 \\ \tilde{f}_2 \end{array} \rr =
 S \lr \begin{array}{c} \tilde{f}_L \\ \tilde{f}_R \end{array} \rr,
\eeq
where the mixing angle $\theta_{\tilde{f}}\in[0;\pi/2]$ is related to the
SUSY-breaking masses $m_{\tilde{L},\tilde{U},\tilde{D}}$ through
\beq
 \tan2\theta_{\tilde{f}}={2m_fm_{LR}\over m_{LL}^2-m_{RR}^2}
\eeq
and the squared mass eigenvalues are given by
\beq
 m_{1,2}^2=m_f^2+{1\over 2}\lr m_{LL}^2+m_{RR}^2\mp\sqrt{(m_{LL}^2-
 m_{RR}^2)^2+4 m_f^2 |m_{LR}|^2}\rr.
\eeq
If these masses are known, a determination of the sfermion mixing angle will
therefore primarily constrain the SUSY and Higgs parameters $A_f$, $\mu$ and
$\tan\beta$.

Since the SUSY partners of the neutral and charged electroweak gauge and Higgs
bosons are all spin-1/2 fermions, they can also mix to form four neutral and
two charged mass eigenstates $\tilde{\chi}^{0}_i$ and $\tilde{\chi}^{\pm}_i$. The
neutralino mass matrix
\bea
 Y &=& \left( \begin{array}{cccc}
  M_1 & 0 &
  -m_Z\,s_W\,c_\beta &
  ~~m_Z\,s_W\,s_\beta \\
  0 & M_2 &
  ~~m_Z\,c_W\,c_\beta &
  -m_Z\,c_W\,s_\beta \\
  -m_Z\,s_W\,c_\beta &
  ~~m_Z\,c_W\,c_\beta &
  0 & -\mu \\
  ~~m_Z\,s_W\,s_\beta &
  -m_Z\,c_W\,s_\beta &
  -\mu & 0
 \end{array} \right)
\eea
and the chargino mass matrix
\bea
 X &=& \left( \begin{array}{c c} M_{2} & m_{W}\, \sqrt{2}\, s_\beta \\
 m_{W}\, \sqrt{2}\, c_\beta &  \mu \end{array}\right)
\eea
depend not only $\mu$ and $\tan\beta$, but also on $M_1$ and $M_2$, the
SUSY-breaking $B$-ino and $W$-ino mass parameters, while $m_Z$ and $m_W$ are the
SM $Z$- and $W$-boson masses and $s_W$ $(c_W)$ is the sine (cosine) of the
electroweak mixing angle $\theta_W$. After electroweak gauge-symmetry breaking and
diagonalization of the mass matrix $Y$ with a unitary rotation matrix $N$, one
obtains the four neutralino mass eigenstates. Their decomposition into gaugino and
higgsino components is of major importance, in particular for the determination of
the nature of dark matter. The application of projection operators leads to
relatively compact analytic expressions for the mass eigenvalues.
Since $X$ is not symmetric, it must be diagonalized by two
unitary matrices $U$ and $V$. Its squared eigenvalues are
\bea
 m_{\tilde{\chi}^\pm_{1,2}}^2 &=& \frac{1}{2}\left\{|M_2|^2+|\mu|^2+
 2 m_W^2 \mp \sqrt{(|M_2|^2+|\mu|^2+ 2 m_W^2)^2 - 4 |\mu M_2- m_W^2
 s_{2\beta}|^2}\right\},
 \label{eq:210}
\eea
and the rotation angle $\theta_+\in[0;\pi]$ of $V$ is uniquely fixed by the two
conditions
\bea
 \tan2\theta_+&=&\frac{2\sqrt{2}m_W \left( M_2^*\,s_\beta + \mu\,c_\beta
 \right)\,e^{i\phi_+}} {|M_2|^2 -|\mu|^2 + 2 m_W^2c_{2\beta}}~~~~{\rm and}\\
 \sin2\theta_+&=&\frac{-2\sqrt{2} m_W \left( M_2^*\,s_\beta + \mu\,c_\beta
 \right)\,e^{i\phi_+}
 }
 {\sqrt{(|M_2|^2 -|\mu|^2 + 2 m_W^2c_{2\beta})^2
 +8 m_W^2\le(M_2^*\, s_\beta + \mu\,c_\beta)\,e^{i\phi_+}\re^2}}.
\eea
If the lightest chargino mass $m_{\tilde{\chi}_1^\pm}$ is known, $|\mu|$ can be
determined as a function of $M_2$ from Eq.\ (\ref{eq:210}). The anomalous magnetic
moment of the muon and rare $B$-decays both favor $\mu>0$. In addition, one often
assumes not only gauge coupling, but also gaugino mass unification at the GUT
scale and therefore $M_1=5/3 \tan^2\theta_W M_2\simeq 0.5 M_2$ at the electroweak
scale, leaving $M_2$ as the only free parameter determining the gaugino/higgsino
decomposition of the neutralinos and charginos.

\section{Correlation with beam polarization}

Our main observation is that the polarization of initial-state quarks,
transmitted through $s$-channel exchanges of weak bosons or $t$- and $u$-channel
exchanges of squarks, can be strongly correlated with the helicity and
gaugino/higgsino mixing of final-state sfermions and neutralinos/charginos.
While the partonic double-spin asymmetry for $s$-channel produced sleptons
\linebreak
$A_{LL} = -1$
is independent of all SUSY parameters, the single-spin asymmetry
$A_L=\d\Delta\sigma_L/\d\sigma$ with
\beq
 \d\Delta\sigma_L = {4 \pi \alpha^2 \over 3 s^2}
 \le u t - m_i^2 m_j^2 \over s^2 \re
 \le - {e_q e_l (L_l + R_l) (L_q - R_q) \over
       8 x_W (1 - x_W) (1 - m_Z^2/s)}
     - {(L_l^2 + R_l^2) (L_q - R_q) (L_q + R_q) \over
       64 x_W^2 (1 - x_W)^2 (1 - m_Z^2/s)^2} \re,
\eeq
$L_l=S_{j1}S_{i1}^*(2T_l^3-2e_lx_W)$, and $R_l=S_{j2}S_{i2}^*(-2e_lx_W)$
is indeed very sensitive to the slepton mixing matrix $S$ and thus the SUSY
parameters, in particular since the squared photon contribution is eliminated.
It is also easier implemented experimentally, {\it e.g.} at the Tevatron, since
protons are much more easily polarized than antiprotons.
Neutralinos and charginos also receive $t$- and $u$-channel contributions from
squarks, which couple mostly to their gaugino components and render the
analysis slightly more complex \cite{Bozzi:2004qq}. However, the single-spin
asymmetry remains also in this case the most interesting observable to determine
the gaugino/higgsino mixing and, e.g., the $W$-ino mass parameter $M_2$.

\section{Experimental prospects}

Polarized protons can be created from atomic or optically pumped ion sources with
a polarization degree $P$ of up to 87\%. During acceleration, this polarization is
partially lost in resonance crossings, but this can be avoided by introducing
Siberian snakes into the storage ring lattice. In 2009, the RHIC accelerator at
BNL has completed a successful $pp$ run with $\sqrt{S}=500$ GeV and $P\simeq50\%$
and accumulated an integrated luminosity $L$ of about 50 pb$^{-1}$, culminating in
the first observation of weak bosons in polarized $pp$ collisions
\cite{Surrow:2010tn}. Several more ten-week polarized $pp$ runs are planned,
aiming for $P=65-70\%$ and $L=266$ pb$^{-1}$ in each of the runs \cite{rhicspin}.
The Tevatron at FNAL is currently operating as an unpolarized $p\bar{p}$ collider
at $\sqrt{S}=1.96$ TeV and has already accumulated more than 4 fb$^{-1}$ in
luminosity. A study
of proton beam polarization, performed in the mid-1990s, proposes the replacement
of some of the dipole magnets to create room for six Siberian snakes, which would
lead again to $P=65-70\%$ \cite{Baiod:1995eu}.
The LHC at CERN will be operating as an unpolarized $pp$ collider at $\sqrt{S}=7$
TeV for two years to reach $L=1$ fb$^{-1}$, then at $\sqrt{S}=14$ TeV to reach
$L=100$ fb$^{-1}$. Further upgrades, focusing on even higher luminosity, are
currently under discussion. It is interesting to remember that the SSC design had
reserved space for Siberian snakes. At the LHC, the
required locations would have to be liberated by some of the dipoles.

\begin{figure}
 \centering
 \includegraphics[width=0.49\textwidth]{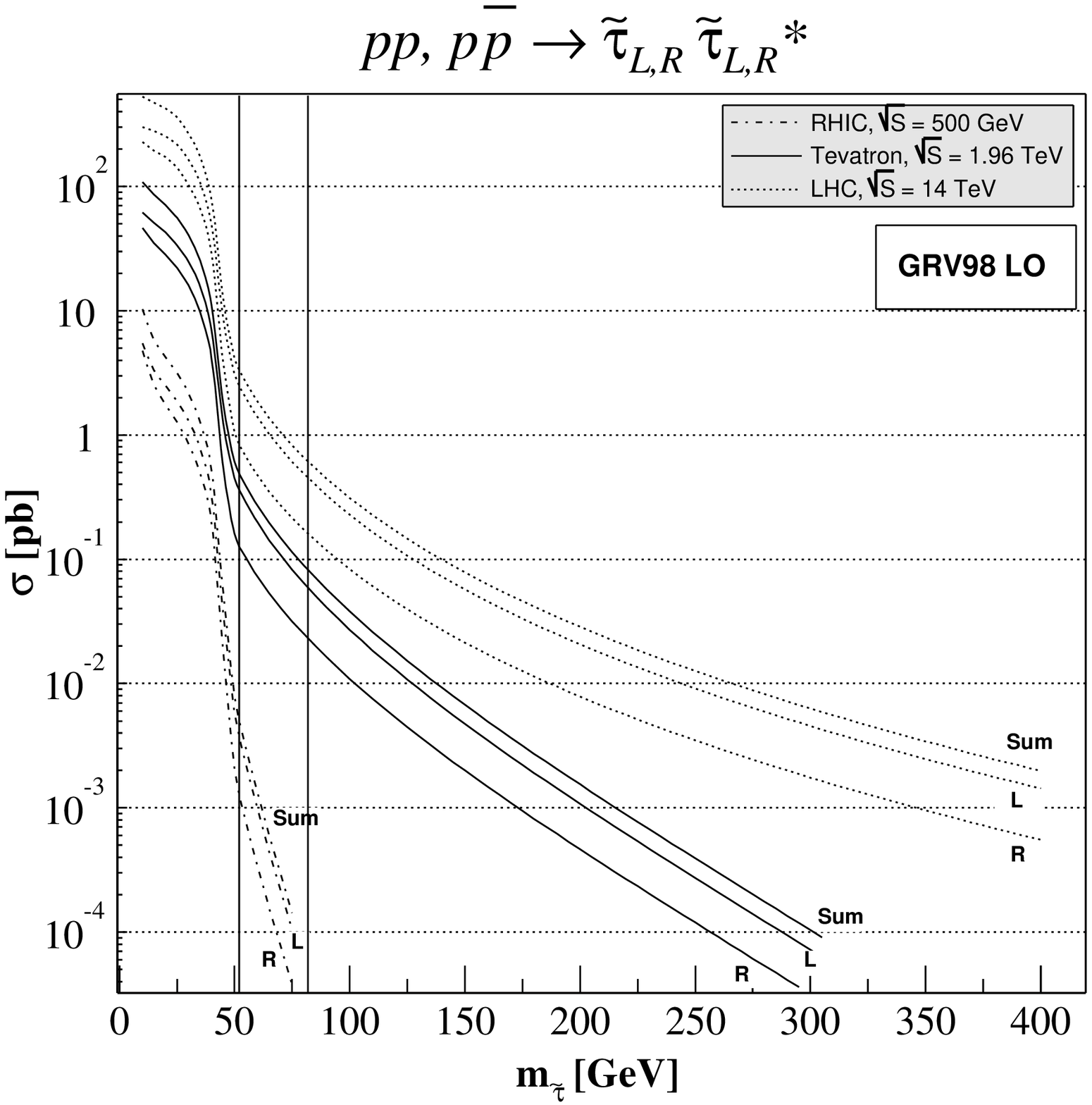}
 \includegraphics[width=0.49\textwidth]{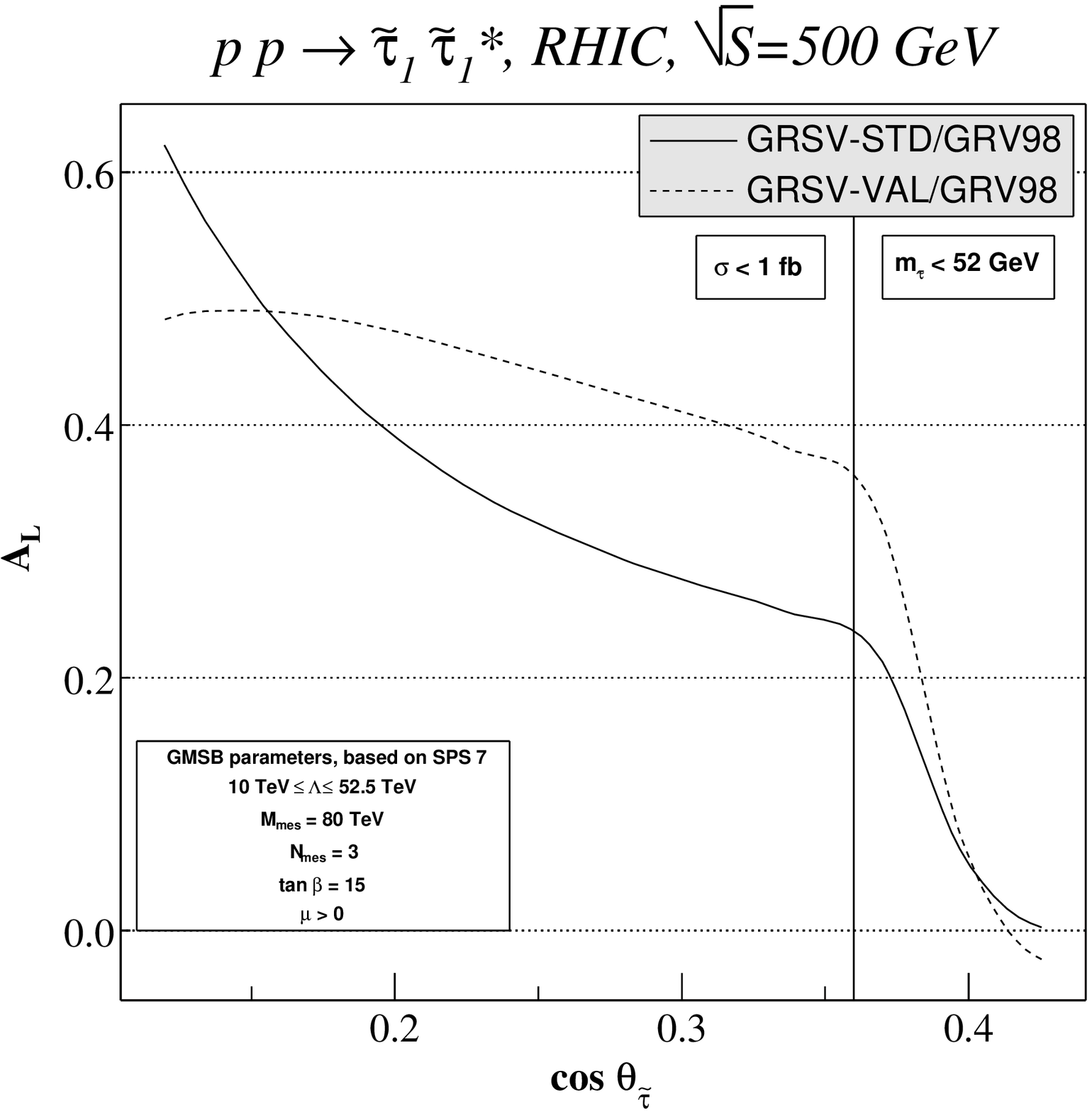}
 \caption{\label{fig:1}Left: Unpolarized hadronic cross sections for pair
 production of non-mixing tau sleptons at the RHIC, Tevatron, and LHC
 colliders as a function of their physical mass. For consistency with the
 polarized cross sections (see below), GRV98 LO parton densities have been
 used. The vertical lines indicate two different stau mass limits of 52
 and 81.9 GeV \cite{Barate:1998zp}. Right:
 Dependence of the longitudinal single-spin asymmetry
 $A_L$ on the cosine of the stau mixing angle for $\tilde{\tau}_1$ pair
 production in a GMSB model at RHIC.}
\end{figure}

In Fig.\ \ref{fig:1} (left) we show the unpolarized hadronic cross sections for
pair production of non-mixing tau sleptons at the RHIC, Tevatron, and LHC
colliders as a function of their physical mass.  Unfortunately, the
observation of tau sleptons, as that of any SUSY particles, will be
difficult at RHIC, which is the only existing polarized hadron collider.
In contrast, tau sleptons will be detectable at the LHC over a large region
of the viable SUSY parameter space up to stau masses of about 400 GeV. At
the Tevatron, the discovery reach extends considerably beyond the current
exclusion limits. For a GMSB model with a light tau slepton, we show the
single-spin asymmetry in Fig.\ \ref{fig:1} (right) as a function of the cosine of
the stau mixing angle. The
asymmetry is quite large and depends strongly on the stau mixing angle.
However, very large values of $\cos\theta_{\tilde{\tau}}$ and stau masses
below 52 GeV may already be excluded by LEP \cite{Barate:1998zp}, while
small values of $\cos\theta_{\tilde{\tau}}$ may be unaccessible at RHIC due
to its limited luminosity, which is not expected to exceed 1 fb$^{-1}$.
Polarization of the proton beam will also not be perfect, and the calculated
asymmetries should be multiplied by the degree of beam polarization $P
\simeq 0.7$. The uncertainty introduced by the polarized parton densities
increases considerably to the left of the plot, where the stau mass 41 GeV
$\leq m_{\tilde{\tau}}\leq$ 156 GeV and the associated values of the parton
momentum fractions become large.

%
%

\begin{figure}
 \centering
 \includegraphics[width=0.49\textwidth]{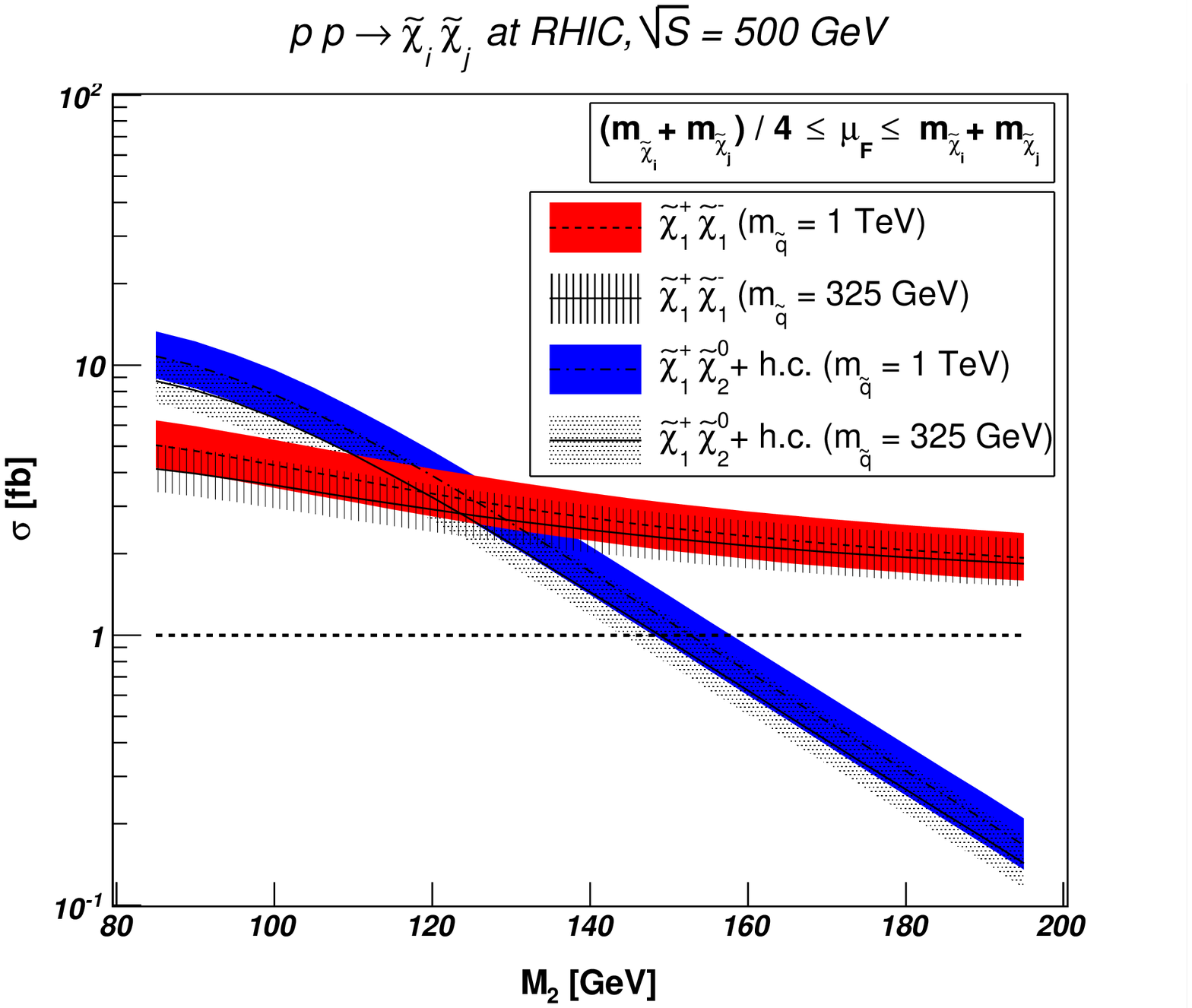}
 \includegraphics[width=0.49\textwidth]{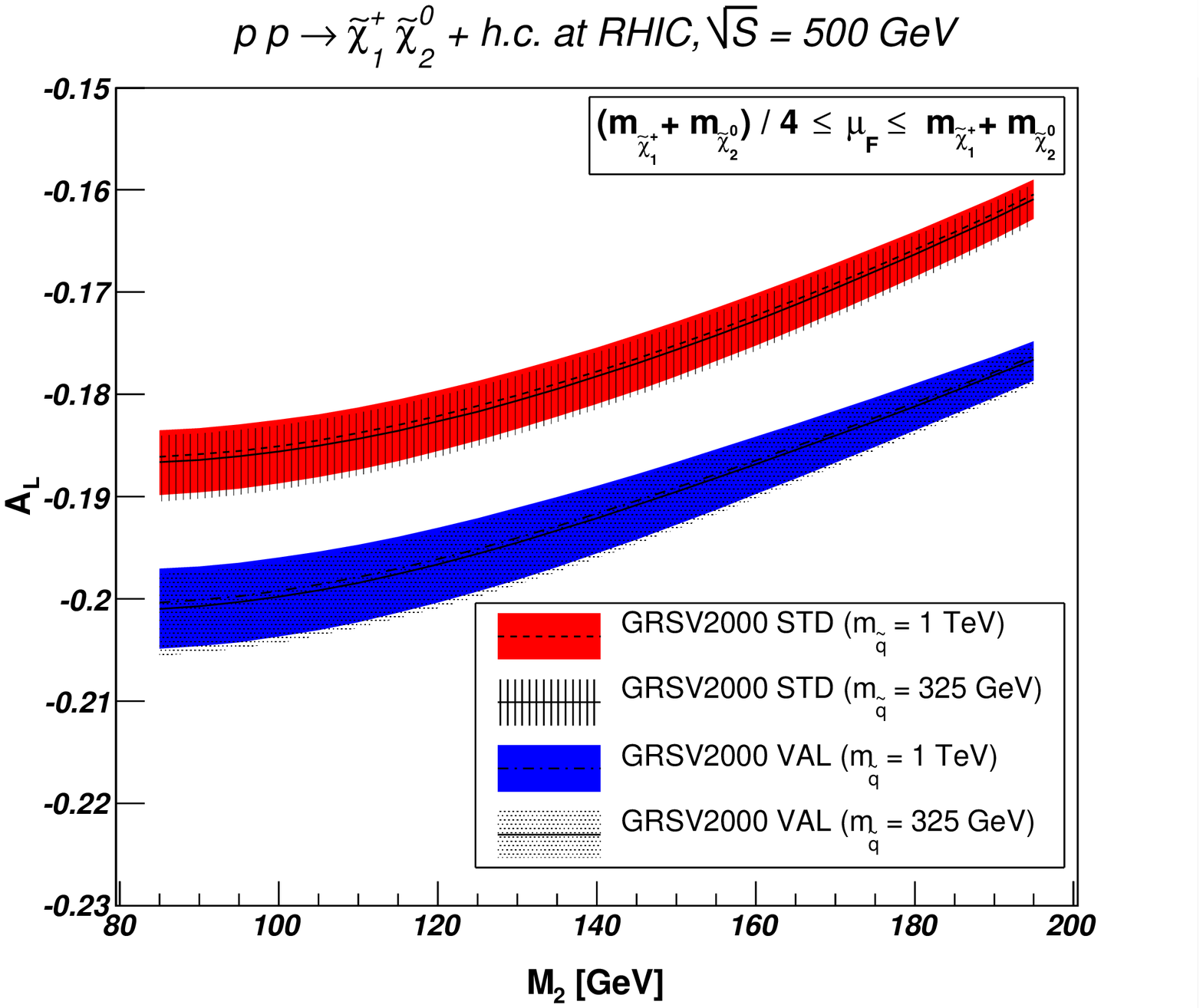}
 \caption{\label{fig:2}Unpolarized gaugino-pair production cross sections
          (left) and single-spin
          asymmetries for chargino-neutralino associated production (right)
          with $m_{\tilde{\chi}^0_2}\simeq m_{\tilde{\chi}^\pm_1}=80$ GeV in
          $pp$ collisions at RHIC and $\sqrt{S}=500$ GeV.}
\end{figure}

In the left part of Fig.\ \ref{fig:2}, we show the total unpolarized cross
%
%
section for the pair production of the lightest chargino of mass 80 GeV
(short-dashed line) and the one for its associated production with the
second-lightest neutralino (dot-dashed line) at the $pp$ collider RHIC.
Both cross sections exceed 1 fb in most of the
$M_2$ range shown and depend little on the squark mass, indicating that
$s$-channel gauge-boson exchanges dominate. We vary the unphysical
factorization scale $\mu_F$ in the
traditional way by a factor of two around the average final state
mass (shaded bands; for more precise predictions see \cite{Bozzi:2006fw,%
Debove:2009ia}).
Among the bosons exchanged in the $s$-channel, the $W$-boson is most
sensitive to the polarization of the initial quarks and antiquarks, and
consequently the single-spin asymmetry for the associated channel, shown in
the right part of Fig.\ \ref{fig:2}, reaches large values of around -20\%.
As the mass of the neutralino increases and the gaugino fractions of the
chargino and neutralino fall up to $M_2\leq200$ GeV, the cross
section and the absolute value of the asymmetry decrease, too.
The uncertainty in the scale variation is with 0.5\% considerably smaller than the
variation in the asymmetry of 2\%, while the uncertainty coming from the polarized
parton densities is with 1.5\% of almost comparable size. Single-spin
asymmetry measurements for associated chargino-neutralino production at the
only existing polarized hadron collider RHIC could therefore be used to
determine the gaugino and higgsino components of charginos and neutralinos,
provided the polarized quark and antiquark densities are slightly better
constrained.
The single- and double-spin asymmetries for neutralino pairs reach similar
sizes as those for the associated channel, since the left- and right-handed
couplings of the $Z$-boson exchanged in the $s$-channel are also different.
However, we do not show them here, since the corresponding cross section is
unfortunately too small at RHIC. The variation of the
asymmetries would, indeed, be quite dramatic: $A_L$ changes its sign from
-20\% to +20\% for $M_2\leq200$ GeV, and $A_{LL}$ falls from -5\% to
-20\%.

\acknowledgments

\noindent
It is a pleasure to thank G.\ Bozzi, J.\ Debove and B.\ Fuks for their
collaboration and the organizers of the DIS 2010 conference for creating a
stimulating atmosphere despite adverse {\em forces majeures}.

\end{document}